\documentclass[a4paper,11pt]{article}
\usepackage{jinstpub} 
\usepackage[nonatbib]{}

\usepackage{array}

\usepackage{amsmath,amssymb,amsfonts}
\usepackage{graphicx}
\usepackage{subcaption}
\usepackage{textcomp,nicefrac}

\begin{document}

\title{Characterization of the response of IHEP-IME LGAD with shallow carbon to Gamma Irradiation}

\author[a,c]{Weiyi Sun}
\author[a,b,1]{Yunyun Fan\note{Corresponding author.}}
\author[a,b]{Mei Zhao}
\author[a,c]{Han Cui}
\author[a,c]{Chengjun Yu}
\author[a,c]{Shuqi Li}
\author[a,c]{Yuan Feng}
\author[a,c]{Xinhui Huang}
\author[a,b,1]{Zhijun Liang}
\author[a,c]{Xuewei Jia}
\author[a,b]{Wei Wang}
\author[a,b]{Tianya Wu}
\author[a,b]{Mengzhao Li}
\author[a]{Jo\~{a}o Guimar\~{a}es da Costa}
\author[d]{and Gaobo Xu}

\affiliation[a]{Institute of High Energy Physics, Chinese Academy of Sciences, \\19B Yuquan Road, Shijingshan, Beijing 100049, China}
\affiliation[b]{State Key Laboratory of Particle Detection and Electronics, \\ Beijing 100049, China}
\affiliation[c]{University of Chinese Academy of Sciences, \\19A Yuquan Road, Shijingshan, Beijing 100049, China}
\affiliation[d]{Institute of Microelectronics, Chinese Academy of Sciences, \\3
Beitucheng West Road, Chaoyang District, Beijing 100029, China}
\emailAdd{fanyy@ihep.ac.cn}
\emailAdd{liangzj@ihep.ac.cn}
\abstract{Low Gain Avalanche Detectors (LGAD) for the High-Granularity Timing Detector (HGTD) are crucial in reducing pileups in the High-Luminosity Large Hadron Collider. Numerous studies have been conducted on the bulk irradiation damage of LGADs. However, few studies have been carried out on the surface irradiation damage of LGAD sensors with shallow carbon implantation. In this paper, the IHEP-IME LGADs with shallow carbon implantation were irradiated up to 2 MGy using gamma irradiation to investigate surface damage. Important characteristic parameters, including leakage currents, breakdown voltage (BV), inter-pad resistances, and capacitances, were tested before and after irradiation. The results showed that the leakage current and BV increased after irradiation, whereas overall inter-pad resistance exhibited minimal change and remained above $10^9\ \Omega$ before and after irradiation.  Capacitance was found to be less than 4.5 pF with a slight decrease in the gain layer depletion voltage (V$_{gl}$) after irradiation. No parameter affected by the inter-pad separation was observed before and after irradiation. All characteristic parameters meet the requirements of HGTD, and this design can be used to further optimization. }

\keywords{Solid state detectors,Radiation-hard detectors,Si microstrip and pad detectors}
\maketitle

\section{Introduction}
\label{sec:introduction}
The Phase-II upgrade of the Large Hadron Collider (LHC), known as the High Luminosity LHC (HL-LHC), aims to achieve a luminosity of $5 \times 10^{34}$ cm$^{-2}$s$^{-1}$, representing a ten-fold increase compared to the LHC\cite{aberle2020submitter, Apollinari2015, Cooke2022}. This higher luminosity leads to significant pile-up effects, with an average of approximately 200 events, necessitating the implementation of a High-Granularity Timing Detector (HGTD) in the forward region of the ATLAS detector. The HGTD will incorporate a new type of silicon sensor, the Low Gain Avalanche Detector (LGAD), to enable precise timing measurements of charged tracks and aid in pile-up suppression \cite{Collaboration}.

LGAD is a type of silicon sensor that offers moderate internal gain, enabling enhanced signal amplitudes and achieving high time and spatial resolutions of better than 20 picoseconds and micrometers, respectively\cite{Moll2018,Moffat2018,Li2022}. However, in the high-intensity beam and irradiation environments of the HL-LHC, LGADs are susceptible to various detrimental effects that can significantly impact their lifetime and performance \cite{bharthuar2021effect,Moll2018,Yang2022}. To ensure the required irradiation hardness, the ATLAS committee has established stringent requirements for the detectors. The LGADs in the HGTD of ATLAS will not be replaced before the completion of half of the data taking of the HL-LHC (2000 fb$^{-1}$). The anticipated total ionizing dose (TID) irradiation dose for the LGADs is expected to be 1.5 MGy\cite{Mazini:2777584}.

Multiple institutes, including BNL, FBK, and NDL, have conducted studies on LGADs, investigating the effects of bulk damage caused by irradiation and observing numerous adverse effects on LGAD performance \cite{Heller2022,Ferrero2019,Tan2021}. However, there has been limited research on surface damage specifically related to LGADs with carbon implantation \cite{hoeferkamp2022characterization}. This study aims to investigate the surface damage induced by gamma-ray irradiation on LGADs with shallow carbon implantation.

To achieve excellent irradiation hardness, the Institute of High Energy Physics (IHEP) collaborated with the Institute of Microelectronics (IME) to develop a new carbon-doped LGAD named IHEP-IME LGAD. This paper focuses on evaluating the performance of the third version of IHEP-IME LGAD (IHEP-IMEv3 LGAD, or v3) under $^{60}$Co gamma-ray irradiation with a maximum dose of 2 MGy. Such irradiation leads to surface damage at the SiO$_2$ and Si-SiO$_2$ interface, inducing oxide charges, interface traps, and point defects in the silicon sensors caused by Compton electrons and photoelectrons \cite{JZhang_2012,Lecoq2020}. The objective of this investigation is to optimize the surface parameters in LGAD design.

This paper is organized as follows: Section \ref{sec:sample} presents the v3 LGAD and describes the sample and fabrication process. Section \ref{sec:irradiation} provides details of the irradiation and test setups. The experimental results are presented and analyzed in Section \ref{sec:result}, followed by a conclusion in Section \ref{sec:conc}.

\section{IHEP-IME LGAD sensors}
\label{sec:sample}
\begin{figure}[t]
\centerline{\includegraphics[width=3.5in]{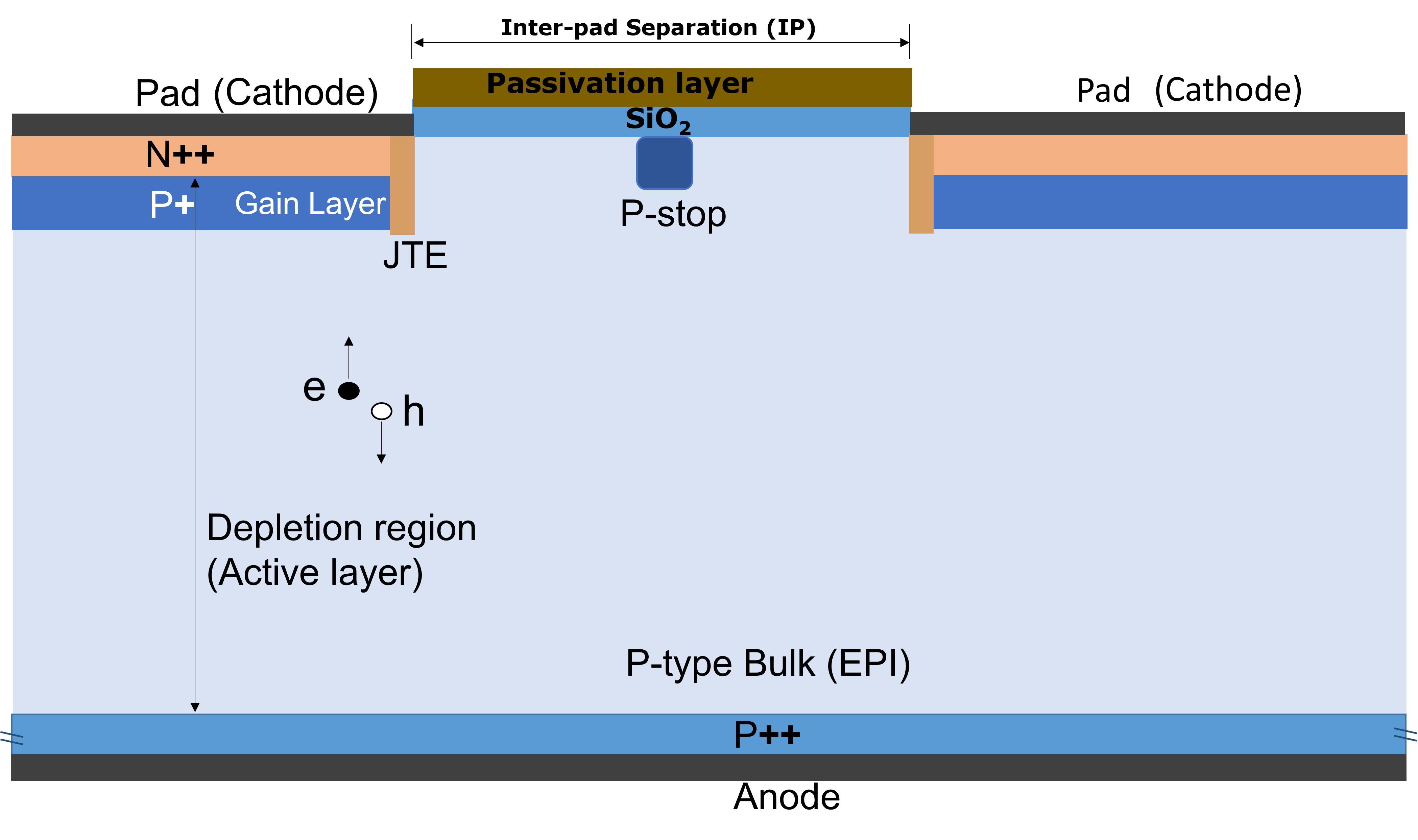}}
\caption{Structure schematic of IHEP-IME LGAD. (Not to scale) }
\label{structure}
\end{figure}

LGAD samples used in this study were mainly taken from wafer 12 of v3. Some LGAD samples of the first version of IHEP-IME LGAD (IHEP-IMEv1 LGAD, or v1) were also taken. The structure of v3 from top to bottom is passivation layer, pad, SiO$_2$, n$^{++}$ layer, p$^{+}$ layer (gain layer), p-type bulk, p$^{++}$ layer and aluminum chuck as illustrated in \figurename~\ref{structure}. In comparison to v3, v1 is also  shallow carbon-doped LGAD, but it has not undergone  surface passivation, which means it does not have the passivation layer shown in \figurename~\ref{structure}. From an external perspective, v1 and v3 appear to be identical.

\begin{figure}[t]
\centerline{\includegraphics[width=3.5in]{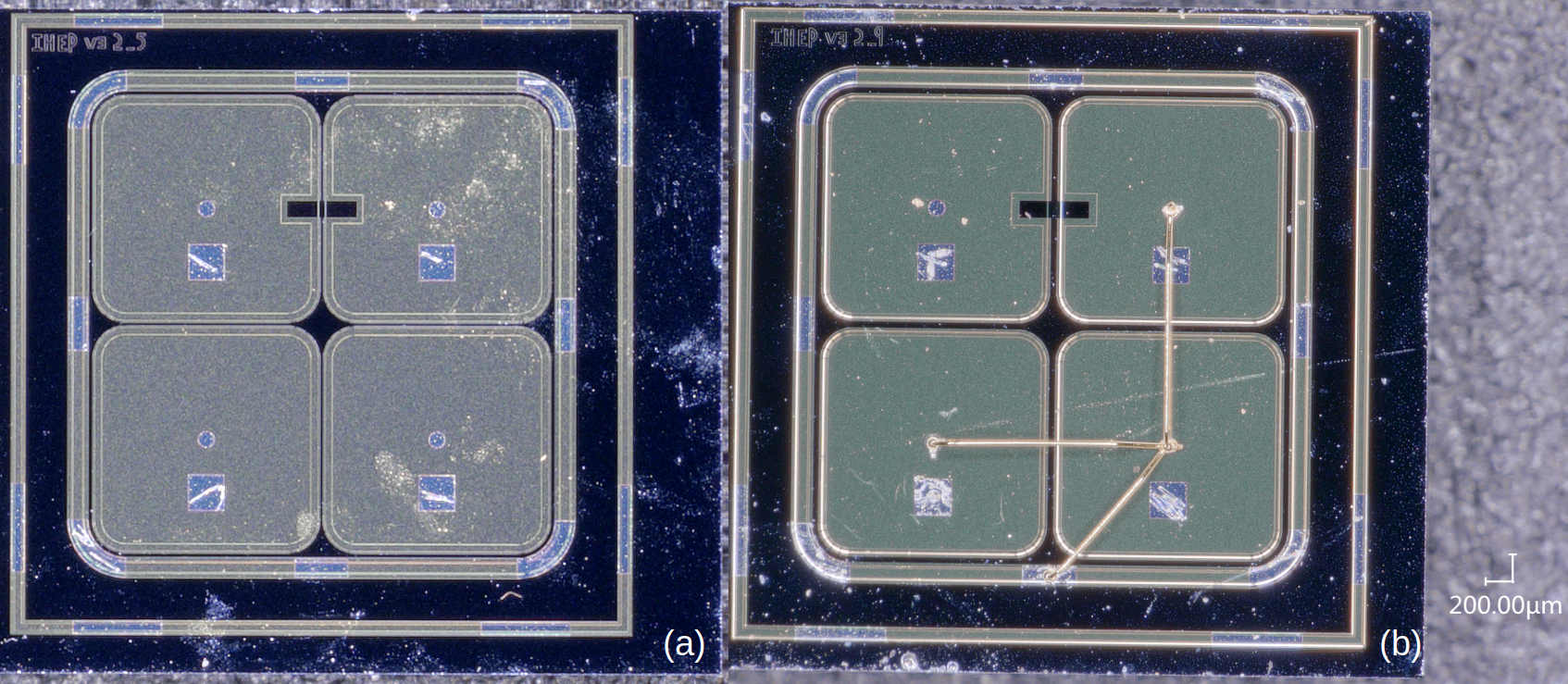}}
\caption{Layout of v3 prototypes 2-5(a) and 2-9(b). Three pads and the guard ring of 2-9 are bonded by gold wires.}
\label{fig1}
\end{figure}

The thickness of p-type bulk is 50 $\mu$m and of p$^{++}$ layer is 725 $\mu$m.  A p-stop and JTEs between n$^{++}$ and p$^{+}$ layers are designed to reduce the lateral current, and all tested LGADs are quad square pad ($2\times 2$ pads, 2 mm $\times$ 2 mm for each pad) structures with different inter-pad separation (IP) from 50 $\mu$m to 100 $\mu$m, surrounded by guard-ring, as summarized in Table~\ref{SIP} and shown in \figurename~\ref{fig1}. Apart from the difference in IP, all other design parameters of the v3 sample are the same.

\begin{table}
\centering
\caption{Sample and its corresponding IP with irradiation dose. The LGAD of v1 and all six types of IP of v3  were irradiated to three doses, namely 10k, 100k, and 2M Gy.}
\label{SIP}
\begin{tabular}{|cc|c|c|c|c|}
\hline
\multicolumn{2}{|c|}{Sample}                     & Irradiation Dose (Gy) & IP ($\mu m$) & Surface Passivation & \multicolumn{1}{l|}{Shallow Carbon} \\ \hline
\multicolumn{1}{|c|}{} & 2-5  & 10k,100k,2M           & 50           & Yes                 & Yes                                 \\ \cline{2-6} 
\multicolumn{1}{|c|}{}                    & 2-6  & 10k,100k,2M           & 60           & Yes                 & Yes                                 \\ \cline{2-6} 
\multicolumn{1}{|c|}{}                    & 2-7  & 10k,100k,2M           & 70           & Yes                 & Yes                                 \\ \cline{2-6} 
\multicolumn{1}{|c|}{v3}                    & 2-8  & 10k,100k,2M           & 80           & Yes                 & Yes                                 \\ \cline{2-6} 
\multicolumn{1}{|c|}{}                    & 2-9  & 10k,100k,2M           & 90           & Yes                 & Yes                                 \\ \cline{2-6} 
\multicolumn{1}{|c|}{}                    & 2-10 & 10k,100k,2M           & 100          & Yes                 & Yes                                 \\ \hline
\multicolumn{1}{|c|}{v1}                  & 2-5  & 10k,100k,2M           & 50           & No                  & Yes                                 \\ \hline
\end{tabular}
\end{table}

The scratches are on the windows of the pads, and the rest is covered by the passivation layer. Since damage to the LGAD caused by $^{60}$Co irradiation is mainly concentrated at the SiO$_2$ surface, different IP designs have different SiO$_2$ areas, and studying the variation pattern among samples of different IP can help optimize the IP design.

The post-flow test results show that the break-down voltage(BV) of the whole wafer is more than 90\% consistent, and therefore, randomly selected samples are representative of the overall situation.

\begin{figure}[t]
\centerline{\includegraphics[width=3.5in]{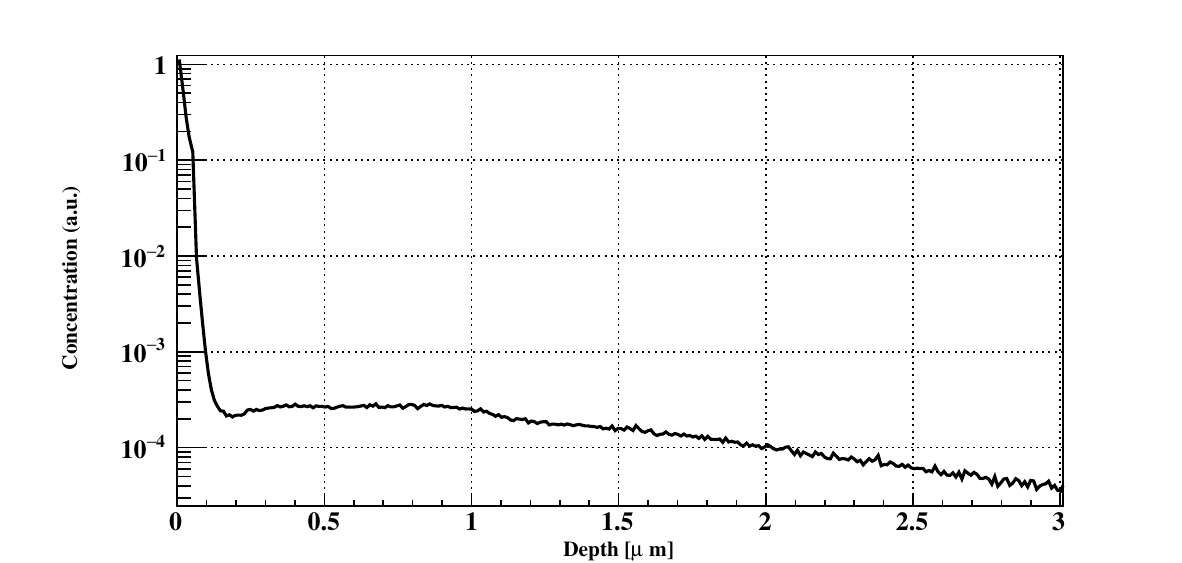}}
\caption{Doping concentration distribution of Carbon, measured through SIMS. }
\label{CDop}
\end{figure}

The wafer was uniformly doped with carbon in order to improve the irradiation hardness, and the doping concentration distribution is measured through Secondary Ion Mass Spectrometry (SIMS). Passivation  SiO$_2$ layers are removed before SIMS is performed. The SIMS result is shown in \figurename~\ref{CDop}. Considering carbon is injected before oxidation and surface passivation, carbon exists in the SiO$_2$ and passivation layers, but the SIMS does not include these two layers.

\section{Irradiation and Experiment Setups}
\label{sec:irradiation}
\subsection{Irradiation}
\begin{figure}[t]
{
\begin{tabular}{c@{}c}
    \begin{subfigure}[b]{.5\columnwidth}
        \centering
        \includegraphics[width=0.5\columnwidth]{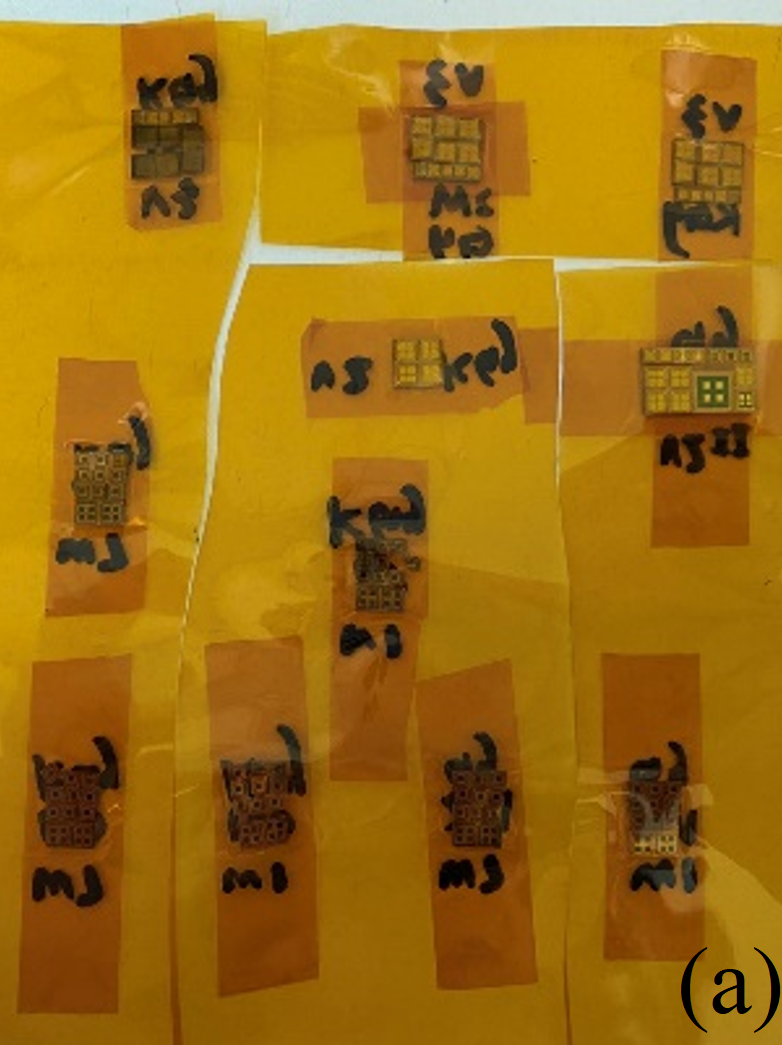}%
    \end{subfigure}&
    \begin{subfigure}[b]{.5\columnwidth}  
        \centering
        \includegraphics[width=0.5\columnwidth]{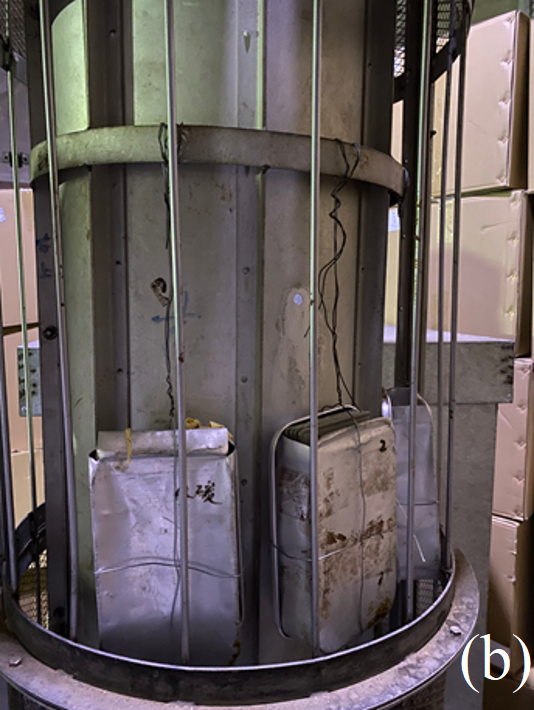}%
    \end{subfigure}
    \end{tabular}
}

\caption{Fixed samples on Kapton tapes (a) are placed in aluminum boxes around the cylindrical  irradiation source (b).}
\label{irr}
\end{figure}

Irradiation test was carried out at the China Institute of Atomic Energy (CIAE). Samples are fixed using Kapton tapes, and placed in the aluminum boxes, around the cylindrical $^{60}$Co irradiation source, which causes damage in silicon sensors by Compton electrons and photoelectrons, as shown in \figurename~\ref{irr}.
Irradiation dose rate is 1.00$\times 10^{4} \pm$15\% Gy/h (~2.77$\pm$15\% Gy/s). The LGAD of all six types of IP was irradiated to three doses, namely 10k, 100k, and 2M Gy, as summarized in Table~\ref{SIP}. Different doses of irradiation for the same IP of LGAD are obtained from different samples from the same wafer. Due to the small sample size compared to the diameter of the radiation source, the irradiation is uniform.

\subsection{Experiment Setups}
All tests, including IV, CV, and Inter-pad resistance, were conducted in the clean room of IHEP at a temperature of 20-22 $ ^\circ$C and a humidity of 10-12\%. The errors of the measuring apparatus are summarized in Table~\ref{eap}.

\begin{table}
\centering
\caption{Errors of experiment apparatus.}
\label{eap}
\setlength{\tabcolsep}{5pt}
\renewcommand{\arraystretch}{1.5}
\begin{tabular}{|c|c|c|}

\hline
Apparatus       & Accuracy   & Used in measurement \\
\hline
Keithley 2400   & 0.012\% & IV CV Resistance    \\
Keithley 2410   & 0.02\%  & IV Resistance       \\
Keysight E4980A & 0.05\%  & CV \\

\hline
\end{tabular}

\end{table}

The leakage currents are measured from $10^{-13}$ A to $10^{-4}$ A, and break-down current is defined as 1 $\mu$A. Negative bias potential is applied to chuck in 2V increment for IV and inter-pad resistances, and in 0.5V increment for CV.

For IV and CV tests, one pad and guard ring are grounded, with the rest pads floating. For inter-pad resistance, three pads and the guard ring are connected by gold wires and grounded, as illustrated in \figurename~\ref{fig1} (b). The rest pad (bias pad) is biased with a voltage range of -1 V to 1 V in 0.5 V increments. The leakage current is measured on ground pads and the bias current is measured on the biased pad. 

Since different experiments require different wiring methods, the effects of different wiring methods with respect to the guard ring plus 0, 1, 2, and 3 pads floating are analyzed. No significant change in leakage current and BV are observed by changing the wiring method, and therefore the following experimental results can be excluded as a result of the test method.

\section{Description and Analysis of Result}
\label{sec:result}
Results of Leakage Current versus Bias Voltage(I-V) and BV are shown in \ref{subsec:IV}; inter-pad resistance is analyzed in \ref{subsec:IpR}; some other issues on bulk damages are discussed in \ref{subsec:CV}.
\subsection{Leakage Current versus Bias Voltage(I-V)}
\label{subsec:IV}

In general, an increase in both leakage current and BV after irradiation was observed. This conclusion holds for all IP. The following is the detailed analysis process.

\begin{figure}[t]
\centerline{\includegraphics[width=3.5in]{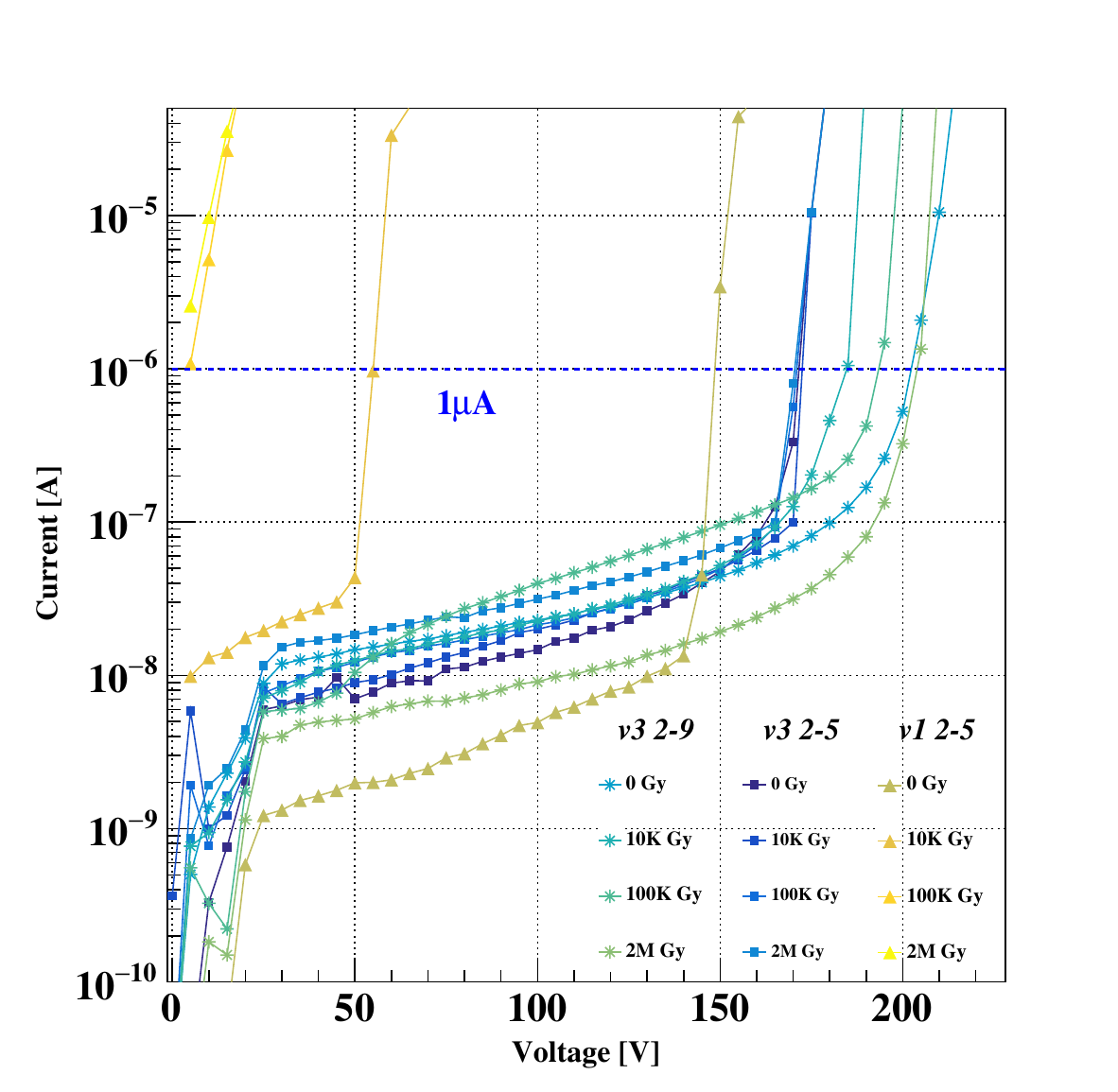}}
\caption{IV characteristic of v3 2-5, 2-9 and v1 2-5. }
\label{fig2}
\end{figure}

I-V characteristics of IHEP-IME LGAD before and after irradiation are measured as shown in \figurename~\ref{fig2}, and take v3 2-5, 2-9, and v1 2-5 as examples.

For v3 2-5, the leakage currents increase with the rise in irradiation dose, reaching three times the pre-irradiation value after exposure to 2 MGy of irradiation. The variation in breakdown voltage (BV) among different irradiation doses is approximately 5 V. Similarly, for v3 2-9, the trend of leakage currents before breakdown resembles that of sample v3 2-5, but BV increases by 20 V after being subjected to 2 MGy of irradiation.
For v1  2-5,  the leakage current is greater than that of v3, and BV is lower than v3 before irradiation. The BV of v1 decreases dramatically after irradiation, indicating that v1 is damaged. v3 samples have undergone a surface passivation process compared to v1, which may be responsible for the increased irradiation hardness but remains to be further investigated. V1 will not be studied in the following due to poor performance.
 
\begin{figure}[t]
\centerline{\includegraphics[width=3.5in]{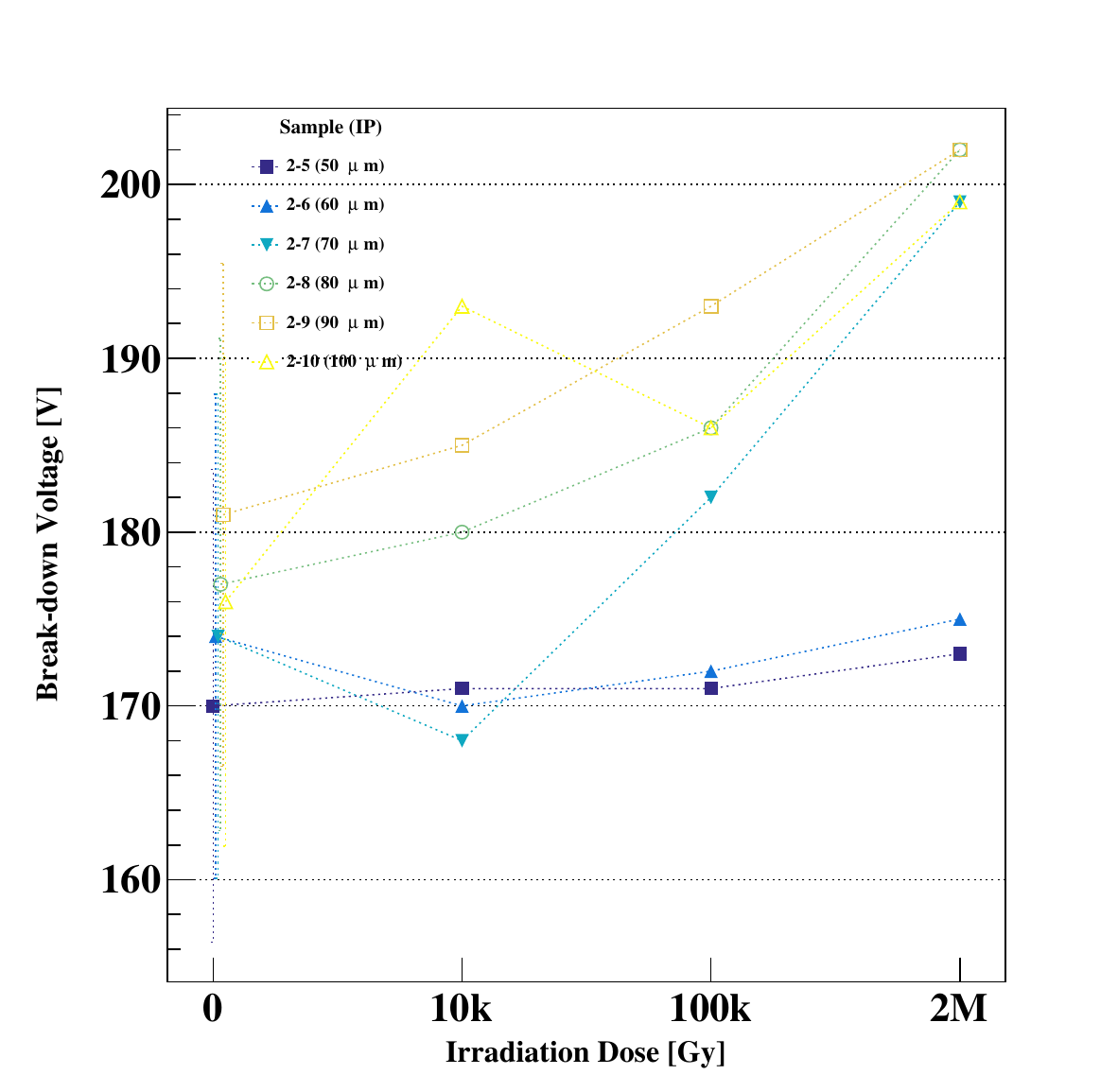}}
\caption{BV of all IP under different irradiation doses. For samples before irradiation, error bars are added according to the spread of BV of the wafer. }
\label{fig3}
\end{figure}

 All samples are investigated in a similar way, and results are shown in \figurename~\ref{fig3}. BV increases with increasing irradiation dose, rising by 5 V-30 V for different samples after 2M Gy irradiation. This trend is particularly significant in samples 2-7, 2-8, 2-9, and 2-10, where the rise is greater than 20 V. Considering different doses of irradiation for the same IP of LGAD are obtained from different samples from the same wafer and the BV of the whole wafer  has about 10\% spread before irradiation, a change of 15V or more in BV after irradiation can be deemed to be caused by irradiation. Consequently, the observed BV rise in samples 2-7, 2-8, 2-9, and 2-10 can be attributed confidently to the irradiation effects. The variations observed in 2-5 and 2-6 are smaller than the 10\% BV spread, and therefore the relationship between BV and irradiation dose cannot be determined in these cases.

\subsection{Inter-pad Resistance versus Bias Voltage}
\label{subsec:IpR}
Inter-pad resistance (R) refers to the resistance between electrodes, specifically the resistance between the grounded and biased pads. In operational LGADs, lateral currents can be generated between electrodes, leading to increased noise and reduced signal-to-noise ratio. A larger R-value helps to reduce these lateral currents. In this study, the R values of v3 samples 2-5, 2-6, 2-7, 2-8, 2-9, and 2-10 were measured.

\begin{figure}[t]
\centerline{\includegraphics[width=3.5in]{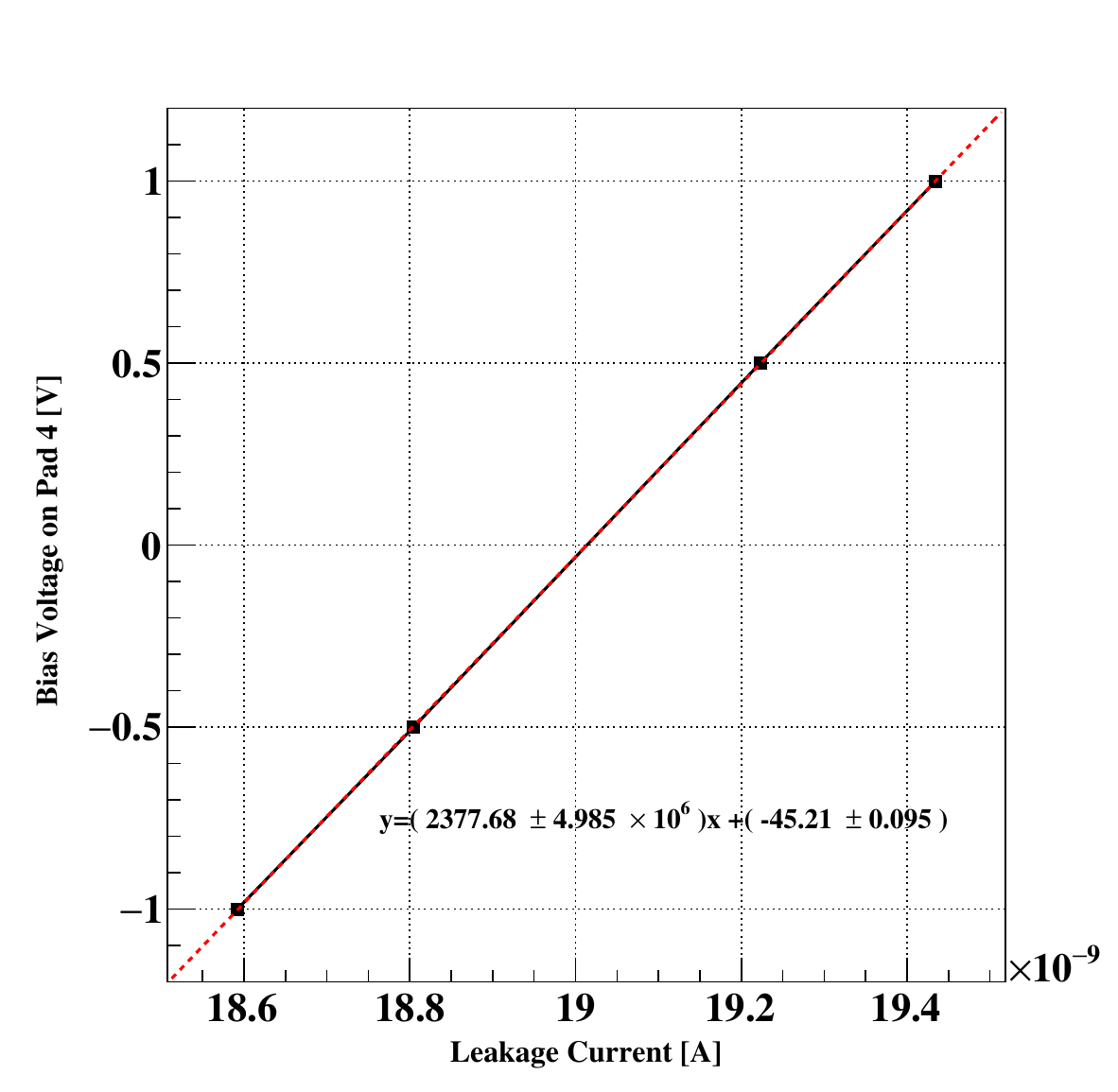}}
\caption{I-V characteristic (solid black line) and linear fit (red dashed line) of v3 2-5 before irradiation at a chuck bias of -85 V.}
\label{fig9}
\end{figure}

To calculate R, a plot of the bias voltage on the biased pad $vs.$ the leakage current is created, and a linear fit is applied to the slope of this I-V characteristic, yielding the resistance value. Figure \ref{fig9} shows the I-V characteristic and its linear fit for v3 2-5 at a chuck bias of -85 V before irradiation. The plot demonstrates excellent linearity with an $R^2$ value greater than 0.99. It is worth noting that the current used to calculate R includes the current from both the substrate and neighbor pads. This can result in a lower calculated resistance value, and the actual resistance is higher than the calculated result. 

\begin{figure}[t]
\centerline{\includegraphics[width=3.5in]{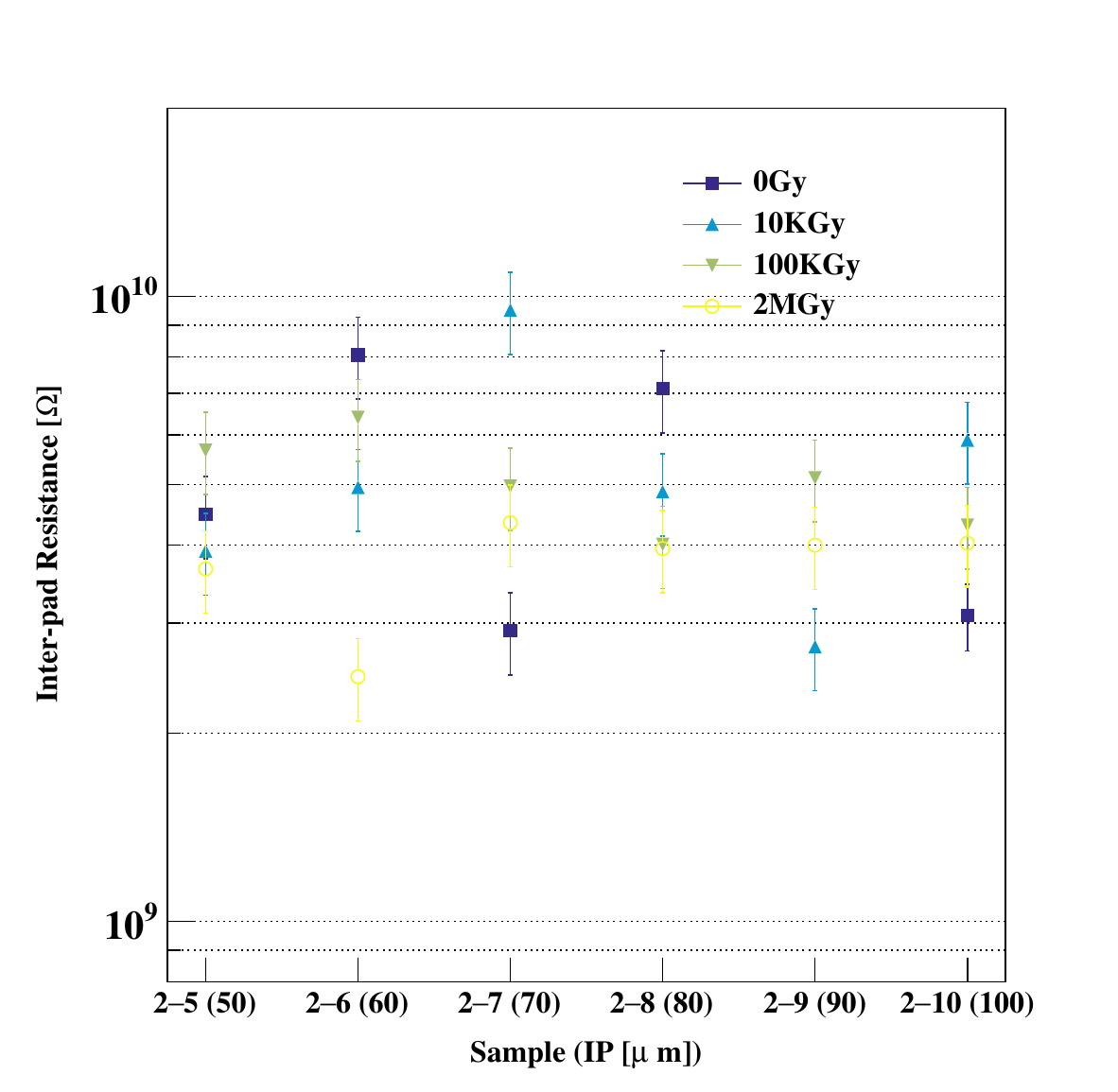}}
\caption{Inter-pad resistance of v3 at chuck bias -85 V of all samples before and after irradiation. }
\label{fig12}
\end{figure}

For this study, the reference operating voltage is selected at 85 V. The R of all samples are shown in Figure \ref{fig12}. All samples, including those irradiated to 2 MGy, exhibit R values greater than $10^9\ \Omega$, which is two orders of magnitude higher than the non-carbon-doped HPK LGAD irradiated to 2.5 MGy, where R decreases by four orders after irradiation \cite{hoeferkamp2022characterization}. HPK's LGAD also features a passivation layer. For v3, no significant trends, such as shifts, increases, or decreases in R, were observed, indicating that irradiation did not have a significant impact on R. Since all samples have R values greater than $10^9\ \Omega$, it can be concluded that the design of v3 meets the design requirements of having R values greater than 1 G$\Omega$.

\subsection{Extra discussion --Capacitance (C-V) and related quantities}
\label{subsec:CV}
\begin{figure}[t]
{
\begin{tabular}{c@{}c}
    \begin{subfigure}[b]{.5\columnwidth}
        \centering
        \includegraphics[width=1.0\columnwidth]{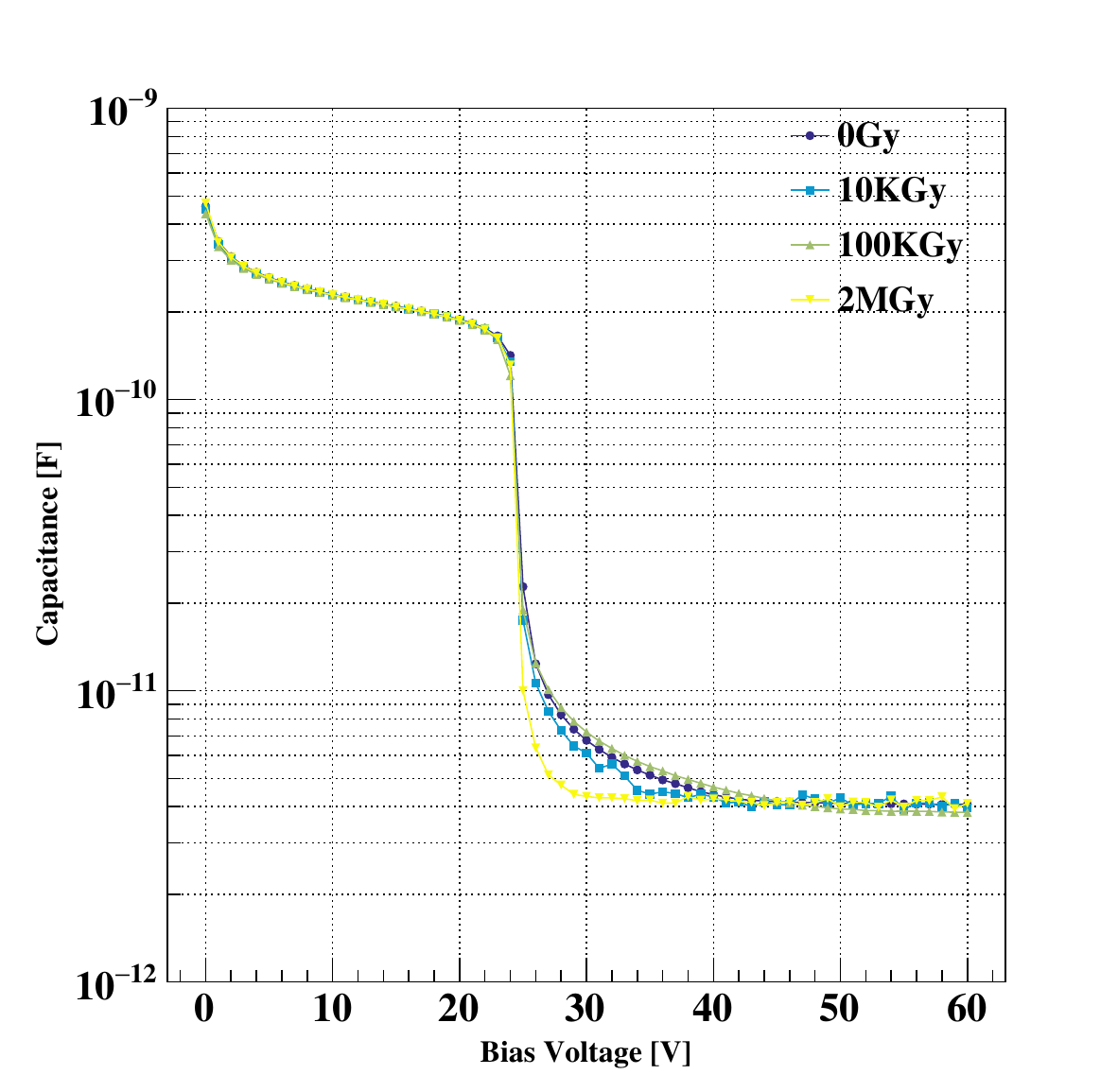}%
        \caption{}
    \end{subfigure}&
    \begin{subfigure}[b]{.5\columnwidth}  
        \centering
        \includegraphics[width=1.0\columnwidth]{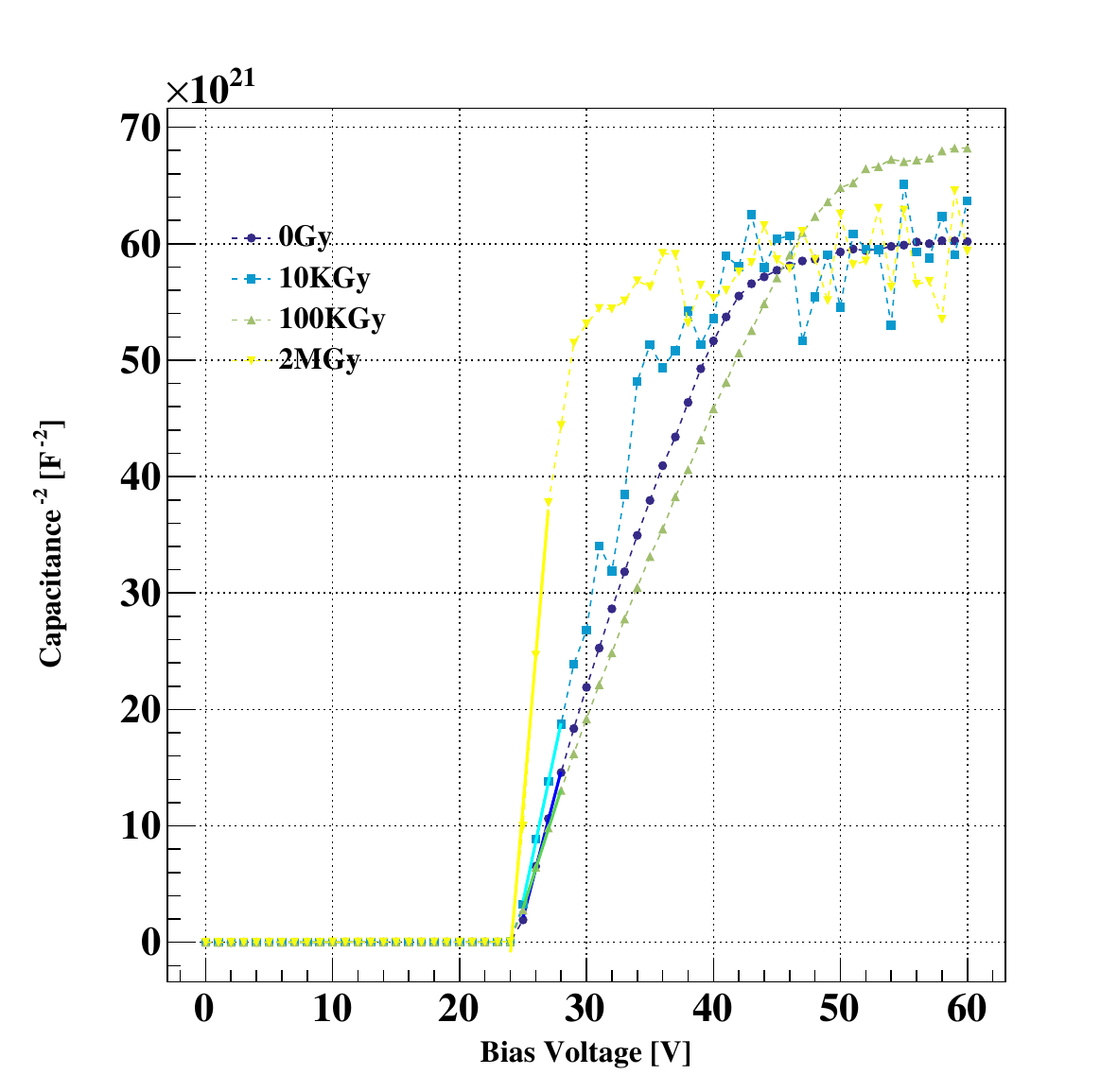}%
        \caption{}
    \end{subfigure}
    \end{tabular}
}

\caption{C-V(a) and $\frac{1}{C^2}$-V (b) characteristics before and after irradiation for v3 2-5 with IP 50 $\mu$m. Data are drawn in dashed lines and fittings in solid in (b). }
\label{fig5}
\end{figure}

This section discusses the capacitance versus bias voltage (CV), V$_{gl}$, acceptor removal constant. 
The CVs of v3 2-5 with IP 50 $\mu$m are shown in \figurename~\ref{fig5}. The figure shows the capacitance values are all less than 4.5 pF, no significant shape change is observed before and after 2M Gy irradiation, and all other samples show the same properties.

Another aspect studied is the gain layer depletion voltage (V$_{gl}$) and the acceptor removal constant, which characterize the bulk damage effect of irradiation.

Calculated V$_{gl} $ is the intersection of the linear fitted line through three points after the turning point and the line y=0 in the $\frac{1}{C^2}$-V curve.

\begin{figure}[t]
\centerline{\includegraphics[width=3.5in]{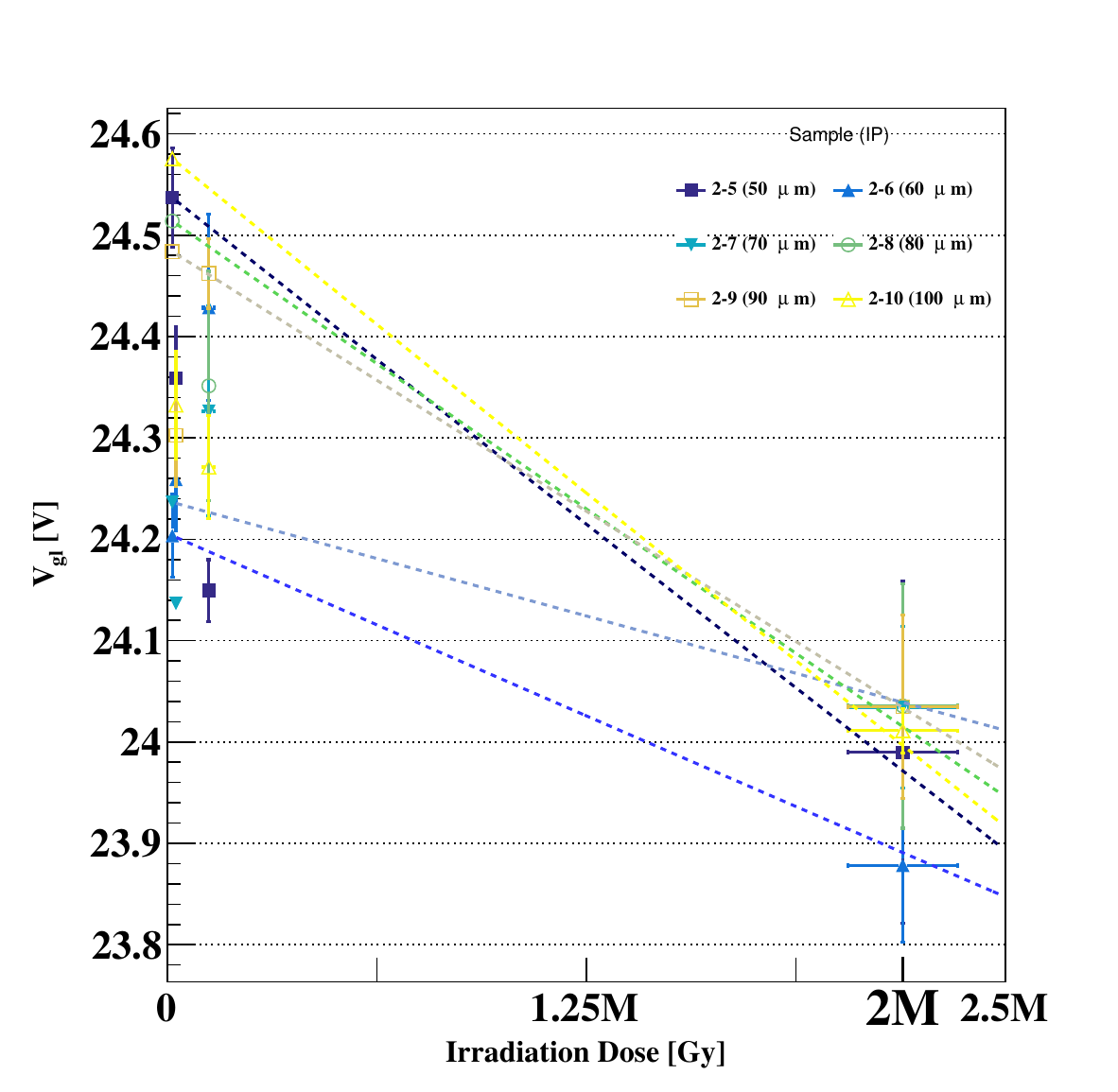}}
\caption{ Fitting of gain layer depletion voltage  V$_{gl}$ as a function of dose to the function ${\rm V}_{gl}={\rm V}_{0,gl}e^{-c\phi}$ for all samples.}
\label{fig6}
\end{figure}

Similarly, V$_{gl}$ is calculated for the other samples and fitted to equation (\ref{eq1}), where V$_{gl}$ and V$_{0,gl}$ represent the gain layer depletion voltages before and after irradiation, $\phi$ is the total ionizing dose measured in Gy, and c is the acceptor removal constant measured in 1/Gy. The fitting results are presented in Figure \ref{fig6}, and the constants c are determined to be in the order of $10^{-9}$ for all samples. It can be observed that there is no significant change in V$_{gl}$ across different IP. This suggests that no bulk damage caused by gamma irradiation is observed.

\begin{equation}
    {\rm V}_{gl}={\rm V}_{0,gl}e^{-c\phi}
    \label{eq1}
\end{equation}

To summarize, except for the slight decrease in V$_{gl}$ after irradiation, no significant effects of irradiation on other parameters are observed.

\section{Conclusion}
\label{sec:conc}
After $^{60}$Co irradiation, the surface components of LGADs were found to be affected, resulting in changes in the studied characteristic parameters. The effects of irradiation on LGADs can be summarized as follows:

\begin{itemize}
\item An increase in leakage current by approximately half an order and a small increase in BV were observed after irradiation.
\item In the case of v3, no significant change in inter-pad resistance was observed before and after irradiation. 
\item There was no correlation between the current used to calculate R parameters and inter-pad spacing, and no significant bulk damage was observed in all tests.
\end{itemize}

Based on these findings, it can be concluded that v3 demonstrates irradiation hardness indicators under the dose of 2M Gy in terms of BV(180-200 V), inter-pad resistance(> 1G$\Omega$), and  capacitance(< 4.5 pF). These results indicate carbon doping combined with surface passivation increases radiation hardness. This study will contribute to the optimization of future LGAD surface designs of inter-pad distance, the gap between the active edge and the guard ring, and the maximum fill factor.

\acknowledgments
This work was supported in part by the National Natural Science Foundation of China under Grant 12105298, Grant 12275290 and Grant 12175252; in part by the State Key Laboratory of Particle Detection and Electronics under Grant No.SKLPDE-ZZ-202315.

\bibliographystyle{plainnat}
\bibliography{ref.bib}
\end{document}